# Joint time and frequency dissemination network over delay-stabilized fiber optic links

Wei Chen[1], Qin Liu[2], Nan Cheng[1], Dan Xu[1], Fei Yang[1], Youzhen Gui[2,3], and Haiwen Cai[1,4]

[1] *Shanghai Key Laboratory of All Solid-State Laser and Applied Techniques, Shanghai Institute of Optics and Fine Mechanics, Chinese Academy of Sciences, Shanghai, 201800, China*
[2] *Key Laboratory for Quantum Optics, Shanghai Institute of Optics and Fine Mechanics, Chinese Academy of Sciences, Shanghai, 201800, China*
[3] *yzgui@siom.ac.cn*
[4] *hwcai@siom.ac.cn*

**Abstract:** A precise fiber-based time and frequency dissemination scheme for multiple users with a tree-like branching topology is proposed. Through this scheme, ultra-stable signals can be easily accessed online anywhere along the fiber without affecting other sites. The scheme is tested through an experiment, in which a modulated frequency signal and a synchronized time signal are transferred to multiple remote sites over a delay-stabilized fiber optic links that are over 50 km long. Results show that the relative stabilities are $5\times10^{-14}$@1 s and $2\times10^{-17}$@$10^4$ s. Meanwhile, compared with each site, time synchronization precision is less than 80 ps. These results can pave the way to practical applications in joint time and frequency dissemination network systems.

## 1. Introduction

Fiber-based frequency peer to peer transfer has been widely discussed and developed with the progress of atomic clocks, wherein Global Positioning System methods are insufficient for short term performance of a microwave or an optical frequency standard [1–3]. Moreover, to extend the range of practical applications, in recent years, considerable interest has been focused on frequency dissemination to multiple users via fiber optic [4–6], which is a challenging task in using a simple control system and an optimized topological cyber structure. As a linear topology, midpoint extraction in a trunk fiber for multiple accesses has been first proposed to address this problem [4–5]. Meanwhile, another tree-like topology, wherein an optical frequency is disseminated to two sites by the multiple reflections compensation method, is also effective [7]. Both schemes produce good results in radio or optical frequency dissemination. To our knowledge, the frequency instability is better than $7\times10^{-14}$@1 s for radio frequency and $8\times10^{-16}$@1 s for optical frequency in the access node [5,8].

However, many applications such as modern large linear accelerators, very long base lines interferometry (VLBI), and deep space networks (DSN) are inclined to have synchronized time with ultra-stable frequency at different remote sites [9–11]. All the aforementioned applications need to determine the exact delay time between each terminal for precise synchronization and to simultaneously access a stable frequency signal to reduce timing uncertainty over free-running time [12]. Consequently, in this study, we propose a scheme to transfer time and frequency jointly to multiple users through a tree-like fiber network with a noise compensation system at each remote site, which will highly reduce system complexity at the local station. Through this scheme, all terminals can be time and frequency synchronized via a delay-stabilized fiber optic link. Relative stabilities of

$5\times10^{-14}$@1s and $2\times10^{-17}$@$10^4$s are obtained while the time synchronization precision of 80 ps is reached.

This paper is organized as follows. The outline of the joint time and frequency dissemination scheme, which exhibits a noise compensation strategy, is first described. Afterward, detailed theories on phase compensation and time synchronization are derived. Then, an experiment is demonstrated and the results of relative frequency stability and synchronization performance are shown by calculating the phase drifting and comparing the absolute time of different sites. Subsequently, some residual factors that may affect the performance of a network system are discussed. Finally, a conclusion of this study is provided and potential applications are presented.

## 2. Experimental scheme

The proposed scheme is illustrated in Figure 1. The local site has a simple structure. No additional components are required and new remote sites can be inserted in the future. The 10 MHz frequency signal from a high-precision clock is multiplied to a higher frequency for the distributed feedback (DFB) laser, which has a considerably lower intensity noise at higher frequency levels that results in higher signal-to-noise ratio (SNR) for compensation [13]. Meanwhile, the one pulse per second (1PPS) time signal is produced by a time generator and locked using the same clock. Dense wavelength division multiplexing technology is applied to multiplex time and frequency signals. Light from two externally modulated lasers with different wavelengths are transferred after multiplexing through the same fiber link. Before it is injected into the long-haul fiber, signals pass through a back-and-forth control structure that consists of an optical coupler (OC), a circulator, an optical filter (OF), and an erbium-doped fiber amplifier (EDFA). This structure is used to pass the forward signal while amplifying and reflecting the backward signal. Meanwhile, the OF is used to stop the backscattering signal injected into EDFA. The joint time and frequency signals along the fiber link can be extracted by an OC. Part of the light will move forward to other terminals, whereas another part may pass through a long-haul fiber to reach a remote site. Given this topological structure, the proposed scheme is considerably more robust than existing schemes for each user [7].

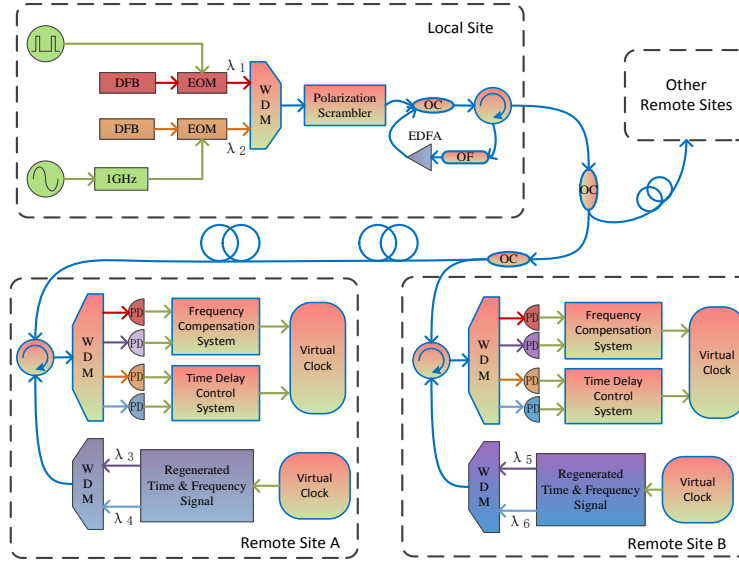

Fig. 1 Schematic of the dissemination structure. DFB: distributed feedback laser, EOM: electro-optic modulator, WDM: wavelength division multiplexer, OC: optical coupler, OF: optical filter, EDFA: erbium-doped fiber amplifier, PD: photodetector

At each remote site, time and frequency signals are detected and recovered by two low-noise photodetectors after they are demultiplexed by a four-channel wavelength division multiplexer (WDM). The use of the other two wavelength channels of this WDM is discussed later. Subsequently, the frequency and time signals are sent to the phase noise compensation system and the time delay control system, respectively. The recovered signal can be regarded as a virtual clock whose signal has an acceptable short-term stability. After a virtual clock is obtained, the basic concept of the noise compensation method can be applied as the traditional round trip method [14–15] at a remote site. In our scheme, two additional wavelength lasers, which are regarded as sensing signals modulated by the regenerated time and frequency signals, are sent back to the local site. After passing through the back-and-forth structure, the sensing signals move along the original path and are demultiplexed by the other two channels of the four-channel WDM at a remote site. Hence, the additional phase noise and time delay induced by the transmitted fiber are detected by sensing signals. They can be respectively compensated by the phase noise compensation and time delay control systems at each remote end.

For additional details, we first focus on the frequency signal, as shown schematically in Figure 2(a). As previously discussed, the 10 MHz frequency signal $V_r$ is boosted to a 1 GHz signal $V_r = \cos(\omega t + \varphi_0)$ without considering its amplitude. Then, the intensity of the 1548.5 nm DFB laser is modulated by an electro-optic Mach–Zehnder modulator. After multiplexing with the time signal, the frequency signal is injected into the long-haul fiber link. At the remote terminal (we use Remote site A as an example, although the same process applies to other remote sites), the frequency signal coming from a local site first passes through an optical delay line, including a temperature controlled fiber optic ring and a fast fiber stretcher as a compensated structure. After being detected by a low-noise photodetector, the frequency signal with one-trip phase drifting can be expressed as $V_{a1} = \cos(\omega t + \varphi_0 + \varphi_p)$. Then, it splits into three parts. In the first part, the phase fluctuation of the final frequency output will be stabilized. In the second part, the output is sent to the phase discriminator as a reference signal. In the third part, the output, as a sensing signal, modulates another DFB laser with a 1547.7 nm wavelength. Given the 100 GHz frequency spacing with the local wavelength, the influences of the backscattered light of the regenerated signal on the frequency signal that is transferred for the first time can be avoided. The optical path is assumed to be symmetrical. Hence, when the sensing frequency signal returns to the remote site, the detected signal after demultiplexing with the corresponding channel becomes $V_{a3} = \cos(\omega t + \varphi_0 + 3\varphi_p)$, which is also sent to the phase discriminator. Therefore, the beat note signal, which is proportional to the phase noise term $2\varphi_p$, is obtained and sent as feedback to the optical delay line for compensation after the proportional-integral-derivative arithmetic is applied. Thus, the final output frequency is stabilized to $V_{a1}' = \cos(\omega t + \varphi_0)$.

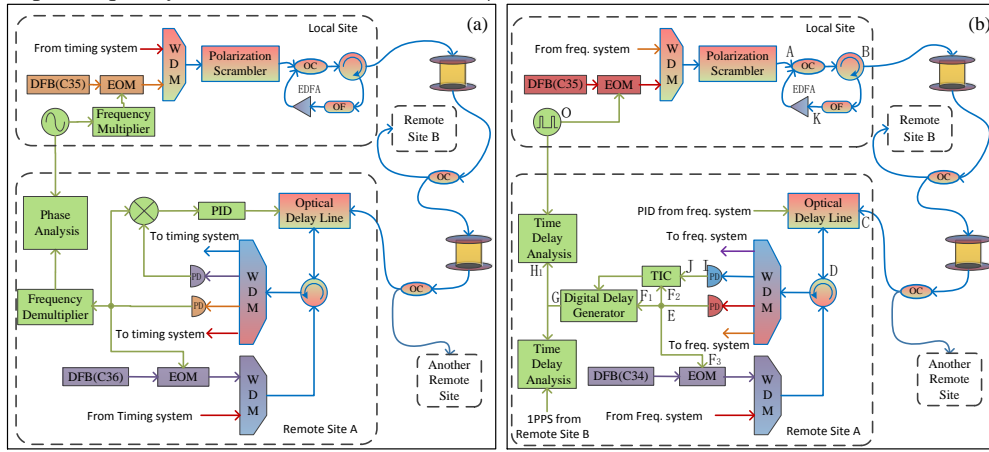

Fig 2. Outlines of the compensation system for time and frequency dissemination for (a) phase noise compensation and (b) time delay synchronization. DFB: distributed feedback laser, EOM:

electron optic modulator, WDM: wavelength division multiplexer, OC: optical coupler, OF: optical filter, EDFA: erbium-doped fiber amplifier, PD: photodetector

Similarly, Figure 2(b) outlines the principle of the time synchronization system that we have developed. For the time signal, more attention is given to the time delay caused by an asymmetrical path. The 1 PPS signal from the time generator is sent to the back-and-forth structure after modulating a 1550.1 nm DFB laser, marked as $t_{OA}$. Point B is the output of the local terminal. The back-and-forth structure has different light path, i.e., $t_{AB} \neq t_{BKA}$. Afterward, the long-haul fiber link to the 1 PPS signal is detected and split into three parts at point E. In the first part, the digital delay generator is triggered. In the second part, the signal is sent to the time interval counter (TIC) as the counting start time. In the third part, the signal is used to modulate another wavelength DFB laser that still has a 100 GHz space with the local one. When the light returns to the remote site, it is detected and sent to the TIC as the counting stop time. The counting time controls the digital delay generator in compensating time delay after calculation. Although the optical path cannot be absolutely symmetrical, particularly at remote sites, all the terminals can still be synchronize after some calibration. That is, the time signal of the remote terminal output (point G) can be always synchronized with the local clock (point O) using the proposed dissemination scheme.

For easy calibration, some cables are selected to have the same length such as $L_{EF_1} = L_{EF_2} = L_{EF_3}$ and $L_{OH_2} = L_{GH_1}$. The intrinsic delay of the system for our synchronization scheme is calibrated in two steps. The calibration process are in the same ambient temperature. In step one, a back-to-back test is performed, in which the local terminal and remote terminal are positioned together without inserting any long-haul fiber link ($t_{BC} = 0$). We need to set a controlled delay time $t_{CDT}$ ($t_{CDT} = t_{F_1G}$ in Figure 2) to ensure remote end synchronization; thus, we obtain

$$t_{OA} + t_{AB} + t_{CD} + t_{DE} + t_{EF_1} + t_{CDT} = 1s. \tag{1}$$

Then, the TIC display is as follows:

$$t_{TIC} = t_{F_3D} + t_{DC} + t_{BKA} + t_{AB} + t_{CD} + t_{DI} + t_{IJ}. \tag{2}$$

In the second step, one or more long-haul fiber is placed between two sites, and thus, the TIC display is changed to

$$t_{TIC'} = t_{F_3D} + t_{DC} + t_{CB} + t_{BA} + t_{AB} + t_{BC} + t_{CD} + t_{DI} + t_{IJ}, \tag{3}$$

where $t_{CB}$ and $t_{BC}$ are the fiber propagation delays in both directions. The propagation delays along the long-haul fiber in both directions are assumed to be equal and equally change with ambient temperature. Thus, compared with those in Equations (2) and (3), the asymmetrical path is dramatically canceled. Propagation delay time is simply expressed as follows:

$$t_{BC} = (t_{TIC'} - t_{TIC})/2. \tag{4}$$

In this status, if the 1 PPS signal is synchronized, then another controlled delay time $t_{CDT'}$ is set. Thus, total delay time is satisfied with the following equation:

$$t_{OA} + t_{AB} + t_{BC} + t_{CD} + t_{DE} + t_{EF_1} + t_{CDT'} = 1s. \tag{5}$$

Using Equations (1) and (4), the controlled delay time is expressed as follows:

$$t_{CDT'} = t_{CDT} - (t_{TIC'} - t_{TIC})/2, \tag{6}$$

which is used to control the digital delay generator. As a matter of fact, the wavelength with a 100 GHz spacing between the back-and-forth light makes the propagation delay time of the two directions different. The effect of chromatic dispersion must be considered, particularly in some long-haul circumstances [16].

## 3. Experimental results

We use two 25 km fiber spools as an example of our proposed fiber networking structure.

Meanwhile, a 10 km and a 1 km fiber optic delay lines are deployed for Remote sites A and B, respectively. The actual optic lengths of the two sites are approximately 60 km and 26 km, respectively. For demonstration, we measure the relative frequency stability of both remote end signals $V_{a1}$ and $V_{b1}$ compared with the frequency reference $V_r$ (with a 5 Hz measurement bandwidth), where both remote sites are located within the same laboratory. The results of the overlapping Allan deviation (ADEV) are shown in Figure 3. Short-time instability can reach $2 \times 10^{-14}$@1 s and $5 \times 10^{-14}$@1 s, respectively, whereas long-time performance can extend to the $10^{-17}$ level at an average time of $10^4$ s. The free-running behaviors of the two sites are similar because they share the same 25 km fiber spools in the same laboratory. Regardless of which remote site, short-time frequency stability does not improve much while the compensation servo is engaged because of the limitations placed by the relative intensity noise of the active device and the frequency noise of the laser carrier. Nevertheless, the long-term effect of temperature changes is dramatically suppressed. For further investigation, we magnify the phase drifting details in the illustration in Figure 3. A slight fluctuation can be detected along time. Two factors should be considered. Phase fluctuations with respect to temperature vary at different wavelengths, and some optical paths are outside the compensation loop. Given that the thermal coefficients of the dispersion is small, i.e., approximately 1.6 fs/(nm·km·K) [17], and the fiber that is outside the loop can be carefully temperature controlled and set as short as possible, the two factors will not affect practical applications.

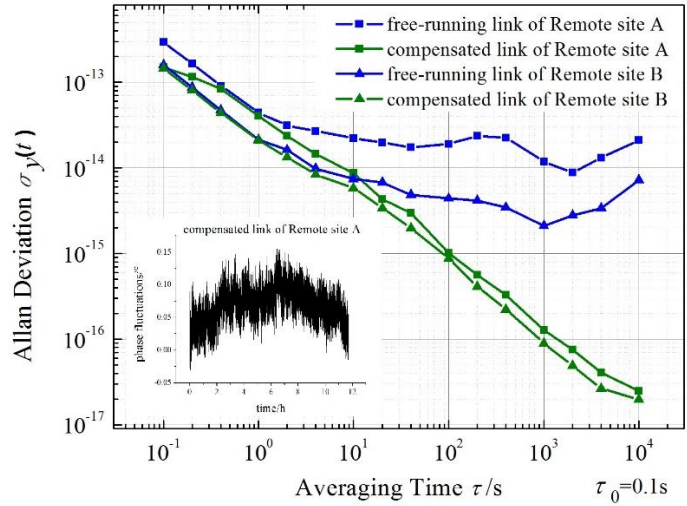

Fig. 3 ADEV of frequency signals at the two remote sites with and without compensation. Inset: phase fluctuations of the compensated link of Remote site A

The exact propagation delay is obtained after two calibration steps. To achieve a more precise calibration, we measure the chromatic dispersion of the two fiber spools using a dispersion analysis machine (MTS-8000, JDSU Corporation, California, USA), which shows that the discrimination of the delay time caused by the dispersion are 836.8 ps/nm and 417.3 ps/nm for Remote sites A and B, respectively, at 1550 nm. Consequently, the propagation delay is revised for synchronization. Hence, fiber length variation is stabilized by the phase noise compensation system and the 1 PPS signal is synchronized after calibration. We use a standard time interval counter (SR620) to compare the absolute delay time of three sets of data: local time with Remote site A, local time with Remote site B, and local time between these two remote sites. The improvement between free-running and compensation links can be observed in corresponding time deviation (TDEV) plot shown in Figure 4. We observe that the TDEV of the compensated link decreases to less than 1.8 ps for the average time between $10^2$–$10^3$ s, but increases if the link is free running. The performance of the synchronization system is illustrated in Figure 5. All signals are nearly stable along time. The time jitter of the 1 PPS signal, which is measured in 10 s intervals, is approximately 7.5 ps (root mean square).The synchronization accuracy values at different sites are 71, 43, and 20 ps. Notably, the time difference between the remote sites is less than the theoretical subtraction of the local site with Remote site A and the local site with Remote site B because of some measurement errors of the TIC. Even so, all remote sites exhibit good synchronization precision that can satisfy the requirements of essentially any practical application at present.

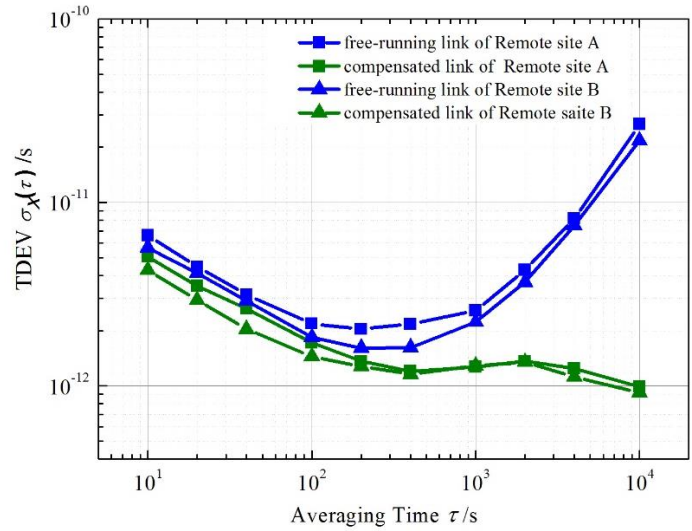

Fig. 4 TEDV of time signals at the two remote sites with and without compensation

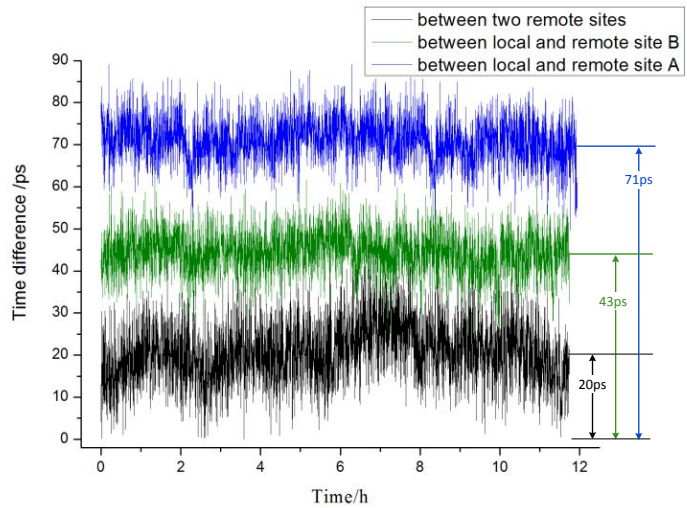

Fig. 5 Time difference between the three sites. Measurements were obtained in 10 s intervals.

## 4. Discussion

The network capacity of the joint time and frequency scheme is limited by two considerable factors. First, the optical power budget should cover the additional optical loss caused by network node insertion. For the entire network design, the splitting ratio of the OC should be assigned according to node distance. Bi-EDFA with high symmetry and low amplified spontaneous emission noise can be inserted into the backbone fiber before the OC. This procedure will provide sufficient optical power without exceeding the threshold of stimulated Brillouin scattering. Thus, additional optical loss can be addressed. Second, wavelength assignment determines the number of the remote terminals in the scheme. The same wavelength is used in the back-and-forth light to sense signals, which causes the problem of backscattering light affecting the SNR of the system. In telecommunication fiber, backscattering light, which is considerably higher than under laboratory conditions, is amplified when it is accompanied by a backward sensing signal. Although a higher frequency can be used to improve carrier-to-noise ratio, it can be a practical drawback of this scheme. By contrast, different wavelengths are designed for each sensing signal to distinguish each remote site. The advantage of this setup is that remote sites will not affect one another. New terminals can be flexibly inserted, even when other terminals are operating. However, this scheme limits the capacity in a single fiber. We use waveband C (1525 nm to 1565 nm) as an example. With one backbone fiber, this waveband can accommodate approximately 25 terminals for a communication channel space of 100 GHz. Furthermore, as the number of terminals increases, the influence of different wavelengths should be considered. If the middle wavelength of waveband C is chosen for local sites, then the maximum wavelength interval between forward time, frequency signals, and backward sensing signals is approximately 20 nm. For frequency signals, the effect of the thermal coefficient of dispersion reaches 32 fs/(km·K). Although the compensation system may be insufficient or excessive, particularly under some long-haul cases, the coefficient is still three orders smaller than the thermal sensitivity of the fiber propagation delay, which is 36.8 ps/(km·K) [18]. For time signals, the delay induced by dispersion may be larger, which is approximately 340 ps/km. However, several dispersion compensation methods can be applied to reduce the effect [19–20]. Furthermore, this delay can be measured by the dispersion analysis machine before calibration. That is, it will not significantly affect time synchronization performance.

## 5. Conclusion

In this study, we propose a scheme to jointly disseminate time and frequency signals to multiple users using an active stabilization and synchronization system. Compared with existing approaches, the proposed scheme satisfies more actual demands and provides for a more flexible and robust online insertion of new remote nodes. Moreover, we perform an experiment to examine the performance of the scheme using a trial link of two 25 km long fiber spools in a laboratory. A time deviation of 1.8 ps@$10^2$ s (for time signals) and Allan deviation of $5\times10^{-14}$@1 s and $2\times10^{-17}$@$10^4$ s (for frequency signals) are obtained, which prove that the scheme is effective. Some contributive factors that may affect system performance are also discussed in this paper. The solutions to these factors will be tested in the next research. Furthermore, the proposed structure can be extended to a branch networking system for fiber-based time and frequency transfer even at a continental scale.

**Acknowledgments**

We would like to thank Professor Zujie Fang for the helpful discussions and guidance that he provided. This work was supported in part by the National Natural Science Foundation of China (No. 61405227).